\documentstyle[pre,aps]{revtex}

   \font\tenmsb=msbm10 scaled\magstep 1
   \font\sevenmsb=msbm7 scaled \magstep 1
   \font\faivemsb=msbm5 scaled \magstep 1
\newfam\msbfam
      \textfont\msbfam=\tenmsb
      \scriptfont\msbfam=\sevenmsb
      \scriptscriptfont\msbfam=\faivemsb
\def\Bbb#1{{\fam\msbfam #1}}
\font\tengothic=eufm10 scaled\magstep 1
\font\sevengothic=eufm7 scaled\magstep 1
\newfam\gothicfam
      \textfont\gothicfam=\tengothic
      \scriptfont\gothicfam=\sevengothic

\newcommand{\Dlt}{\Delta}
\newcommand{\lbd}{\lambda}
\newcommand{\sgm}{\sigma}
\newcommand{\vp}{\varphi}
\newcommand{\ep}{\varepsilon}
\newcommand{\be}{\begin{equation}}
\newcommand{\ee}{\end{equation}}

\begin{document}

\draft

\title{Weighted Fixed Points in Self--Similar Analysis of Time Series}

\author{V. I. Yukalov\footnote{The author to 
whom correspondence is to be addressed}} 

\address{Bogolubov Laboratory of Theoretical Physics\\
Joint Institute for Nuclear Research, Dubna 141980, Russia}

\author{S. Gluzman}
\address{International Center of Condensed Matter Physics\\
University of Brasilia, CP 04513, Brasilia, DF 70919-970, Brazil}

\maketitle

\vskip 2cm

\begin{abstract}

The self--similar analysis of time series is generalized by introducing 
the notion of scenario probabilities. This makes it possible to 
give a complete statistical description for the forecast spectrum by 
defining the average forecast as a weighted fixed point and by 
calculating the corresponding a priori standard deviation and variance 
coefficient. Several examples of stock--market time series illustrate the 
method.

\end{abstract}

\newpage

\section{Introduction}

Time series analysis and forecasting have a long history and abundant 
literature, to mention just a few Refs. [1--5]. When analysing time 
series, one usually aims at constructing a particular model that could 
represent the available historical data and, after such a model is 
defined, one could use it for predicting future. This kind of approach 
has been found to be rather reasonable for describing sufficiently stable 
evolution, but it fails in treating large fluctuations like those 
happening in stock markets. There is a growing understanding that this 
failure is caused by the principal inability to take into account, with any 
given model, quite irregular evolution of markets whose calm at large 
development is occasionally interrupted by sudden strong deviations resulting 
in booms and crashes [6]. Such abrupt changes are not regular cyclic 
oscillations [7] but rather are chaotic events, alike heterophase fluctuations
in statistical systems [8]. Similarly to the latter, strong market 
fluctuations are also of coherent nature, having their origin in the 
collective interactions of many trading agents. The coherent collective 
behaviour of traders is often termed the crowd or herd behaviour [9--11], 
which ascribes a negative meaning to this, although one should remember that 
the process of price formation through the market mechanism is always 
collective. The motion of stock markets is essentially nonlinear and 
nonequilibrium, which makes them one of the most complex systems existing 
in nature, comparable with human brain. Market crashes are somewhat analogous 
to critical phenomena in physical systems [12--14], with the precursor 
signals, reminding heterophase fluctuations [8], being manifested as specific 
log-periodic oscillations [15,16]. To our understanding, a market is a 
nonequilibrium system where two trends, bearish and boolish, are competing. 
This competition results sometimes in random fluctuations all of which by 
their nature are similar to heterophase fluctuations. But the largest among 
them are called crashes or booms, while the smaller ones, usually accompanying
the large fluctuations, are termed precursors and aftershocks, depending on the
time of their occurrence with respect to the main large fluctuations. A market
crash can also be compared with an avalanche transition between two different 
metastable states, as it happens in random polymers [17], or with spinodal 
decomposition. However, these analogies are only qualitative and do not allow 
one to straightforwardly extend the methods of statistical physics to the 
quantitative description of markets.

A novel approach to analysing and forecasting time series has been 
recently suggested [18--20]. This technique, being based on the 
self-similar approximation theory [21--29] can be called the {\it 
self-similar analysis of time series}. In this approach, instead of 
trying to construct a particular model imitating the dynamical system 
generating time series, we assume that the evolution of the system is on 
average self-similar. This is the same as to say that the dynamics 
of the considered system is predominantly governed by its own internal 
laws, with external noise being a small perturbation. Since the observed 
time-series data are the product of such a self-governed evolution, the 
information on some kind of self-similarity is hidden in these data. The 
role of the self-similar analysis is to extract this hidden information. 
The way of doing this has been advanced in our earlier works [18--20] 
where, however, there were missing an important point related to the 
intrinsically probabilistic nature of any forecast. Really, the arbitrage 
opportunities, even assumed as practically riskless, have to be represented 
by probabilities [30]. So that, for instance, a crash is not a certain 
deterministic outcome of a bubble, the date of the crash being random [13], 
the magnitude of a crash being also a random variable. Thus, the problem we 
need to solve is how to construct a priori probabilities characterizing the 
spectrum of possible forecasts in the frame of the self-similar analysis 
[18--20]. When one tries to model the stochastic process, whose realization 
is a time series, by a system of stochastic equations, then one often can 
find the related probabilities as a solution of a Fokker-Planck-type equation. 
Although this problem is not as trivial even for seemingly simple linear 
stochastic processes which, in the case of multiplicative noise, can 
exhibit rather unexpected behaviour with large intermittent bursts [32]. 
And the problem of dealing with nonlinear stochastic equations is 
incomparably more complicated. Moreover, some people advance the 
following principal objection against the belief that all random 
processes, including those related to markets, can be modelled by 
stochastic differential or difference equations. One tells that it 
is only relatively stable recurring processes, like seasonal variations, 
can be successfully modelled by particular equations. Contrary to this, 
such intricate organisms as stock markets cannot, because of their 
extreme compexity, be described over substantially long period of time 
by any system of concrete stochastic equations.

However we do not think that stock markets, as any other statistical 
ensemble of interacting agents, is completely random and absolutely 
unpredictable. But rather, as any other complex organism, markets do 
posses some basic self-similar trends, the information on which is hidden 
in the past data. The aim of analysing time series should be in 
extracting the hidden information about the basic tendencies of the 
process, whose knowledge would make it possible to forecast at least the 
near future. As far as the analyzed time series is usually a realization 
of a random process, it would be naive to expect that it is always 
feasible to predict everything for sure. Certainly not! But what could be 
possible, and what would be the main aim of analysis, is to present a 
spectrum of admissible forecasts weighted with the corresponding 
probabilities. In other words, the outcome of an analysis must be not 
just one number but a set of possible scenarios with the probabilities 
assessing the related risks.

In the present paper we make the necessary step in developing the 
self-similar analysis [18--20] by organizing it in the truly statistical 
form. We define the probabilities of different scenarios and show how the 
method works considering several time series. As examples, we choose 
market time series that are the most difficult case. And among them, we 
select the events accompanied by the rise and blowing up of the so-called 
bubbles, since such nonmonotonic cases are the most hard for description.

A time-series bubble is an event corresponding to a fast rise of the 
time-series values, which abruptly changes to a burst, that is to a 
sudden drop of the values, during the time of order of the time-series 
resolution. In general, bubbles are universal and happen in various time 
series. The time-series bubbles are mostly discussed in connection with 
markets, being for them very common and for many participants quite dramatic. 
Keeping in mind pictures representing time series, one may talk about the 
{\it bubble temporal structures}.

\section{Statistical Self-Similar Analysis}

A time series is an ordered sequence 
$\Bbb{X}\equiv\{ x_n|\; n=0,1,2,\ldots\}$ which is a representation of a 
stochastic process with discrete time $t=0,1,2,\ldots$. A given set 
$\Bbb{X}_N\equiv\{ x_n|\; n=0,1,2,\ldots,N\}$ of $N+1$ elements 
representing historical data can be called the {\it data base}. The 
problem we consider is how, with the given data base $\Bbb{X}_N$, to 
predict the value $x_{N+\Dlt t}$ that would occur at a later time 
$t=N+\Dlt t$. That is, forecasting is a sort of an extrapolation procedure
for stochastic processes.

Let us define the triangle family of subsets of the data base $\Bbb{X}_N$ 
in the following way:
$$
{\bf\Phi}_0 \equiv \{\vp_{00} = x_N \} \; ,
$$
$$
{\bf\Phi}_1 \equiv \{\vp_{10} = x_{N-1}\; , \; \vp_{11}=x_N \} \; ,
$$
\be
{\bf\Phi}_2 \equiv \{\vp_{20} = x_{N-2}\; , \; \vp_{21}=x_{N-1}\; ,
\; \vp_{22}=x_N \} \; ,
\ee
$$
........................................................
$$
$$
{\bf\Phi}_N \equiv \{\vp_{N0} = x_0\; , \; \vp_{N1}=x_1\; , \;\ldots\; ,
\; \vp_{NN}=x_N \} = \Bbb{X}_N \; .
$$
The sequence $\{ {\bf\Phi}_k\}_{k=0}^N$ of the subsets ${\bf\Phi}_k$ forms a 
tower since
$$
{\bf\Phi}_k \subset {\bf\Phi}_{k+1} \qquad (k=0,1,2,\ldots,N-1) \; .
$$
The ordered family (1) will be termed the {\it data tower}.

For each member ${\bf\Phi}_k$ of the data tower (1), we introduce a 
polynomial function
\be
f_k(t) \equiv \sum_{n=0}^k \; a_n\; t^n \qquad (0\leq t\leq k)
\ee
of a continuous variable $t$, with the coefficients $a_n$ defined by the 
algebraic system of equations
\be
f_k(n) =\vp_{kn} \qquad (n=0,1,2,\ldots,k) \; .
\ee
This polynomial function uniquely represents the data, from 
$x_{N-k}=\vp_{k0}$ to $x_N = \vp_{kk}$ pertaining to the subset 
${\bf\Phi}_k$. Then, predicting the values $x_{N+\Dlt t}$ of the time 
series $\Bbb{X}$ is equivalent to the extrapolation of the function (2) 
to the region $t>k$.

As a tool for extrapolation we employ the self-similar exponential  
approximants [32]. To this end, starting with a polynomial function (2), 
we construct the nested exponential
\be
F_k(t,\tau) = a_0 \exp\left ( \frac{a_1}{a_0}\; t \exp\left (
\frac{a_2}{a_1}\; t \ldots \exp \left ( \frac{a_k}{a_{k-1}}\; \tau\; t
\right )\right )\ldots \right ) \; ,
\ee
in which $\tau\geq 0$ is a control function playing the role of the 
minimal time necessary for reaching a fixed point. This control function 
$\tau$ will be called the {\it control time}. It is convenient here 
to use the fixed-point equation in the form of the minimal-difference 
condition [33]
\be
F_k(t,\tau) - F_{k-1}(t,\tau) = 0 \qquad (k\geq 2) \; .
\ee
This equation defines the control time $\tau_k(t)$ as a function of $t$ 
for $k\geq 2$. For $k=1$, we put $\tau_1\equiv 1$. With the form (4), 
equation (5) results in the equation
\be
\tau = \exp\left ( \frac{a_k}{a_{k-1}}\; t\; \tau\right ) \; .
\ee
When $a_k/a_{k-1}\leq 0$, then Eq. (6) always possesses one real solution 
$\tau_k(t)$. But when $a_k/a_{k-1} > 0$, there may be one, two, or no 
real solutions. If we have two real solutions, we need to select the 
minimal of them, remembering that $\tau$ is, by definition, the 
minimal time necessary for reaching a fixed point. If Eq. (6) has no real 
solutions, two ways are admissible. One would be to look for a minimum of 
the difference $|F_k-F_{k-1}|$, instead of accepting Eq. (5). Another 
way, when there is no exact solution of Eq. (6), is to define an 
approximate solution to Eq. (5) by iterating the latter as follows:
$$
F_k(t,\tau) = F_{k-1}(t,\tau_{k-1}) \; ,
$$
which, under the known $\tau_{k-1}(t)$, defines an approximate value for 
$\tau_k(t)$. After the control time is found, substituting it in the 
nested exponential (4), we obtain the self-similar approximant
\be
f_k^*(t) \equiv F_k(t,\tau_k(t)) \; .
\ee
This form can be used for extrapolating the polynomial function (2) to 
times $t>k$.

Thus, with a given data base $\Bbb{X}_N$, we can construct a spectrum
$\{ f_k^*(t)\}$ of $N$ different forecasts suggesting different scenarios 
for the future behaviour of the time series considered. How could we 
characterize the probabilities of these scenarios? The answer to this 
question can be done by invoking the stability analysis [23--25].

Define the function $t_k(\vp)$ by the equation
$$
F_1(t,\tau_k(t)) = \vp \; , \qquad t = t_k(\vp) \; .
$$
Substituting $t_k(\vp)$ into Eq. (7), we get
$$
y_k^*(\vp) \equiv f_k^*(t_k(\vp)) \; .
$$ 
The family of endomorphisms $\{ y_k^*\}$ can be considered as a cascade 
whose trajectory $\{ y_k^*(\vp)\}$ is, by construction, bijective to the 
approximation sequence $\{ f_k^*(t)\}$. For this approximation cascade, 
we may define the local multipliers
\be
\mu_k^*(\vp) \equiv \frac{\partial}{\partial\vp} y_k^*(\vp) \; ,
\ee
whose images in time are given by
\be
m_k^*(t) \equiv \mu_k^*( F_1(t,\tau_k(t))) \; .
\ee
Recall that here and in everywhere what follows $k\geq 1$. The local 
multiplier (9) can be presented in the form of the variational derivative
\be
m_k^*(t) = \frac{\delta F_k(t,\tau_k(t))}{\delta F_1(t,\tau_k(t))} \; ,
\ee
which suggests a convenient for practical purposes expression
\be
m_k^*(t) =
\frac{d}{dt} F_k(t,\tau_k(t)){\Large /} \frac{d}{dt} F_1(t,\tau_k(t))\; .
\ee
From these definitions, it follows that $m_1^*=1$, while from the 
fixed-point condition (5), one has $m_2^*=1$. So that we always have
$$
m_1^*(t) = m_2^*(t) = 1 \; .
$$
The cascade trajectory at the time $k$ is stable provided that
\be
|m_k^*(t)| \leq 1\; ,
\ee
where the equality stands for the neutral stability. It looks natural to 
assume that the most probable scenario corresponds to the most stable 
point of the cascade trajectory, that is to the point $k$ where $|m_k^*|$ 
is minimal. When we are interested in the prediction for the time 
$t=k+\Dlt t$, the most stable point $k=k^*$ is given by the condition
\be
\min_k | m_k^*(k+\Dlt t)| \rightarrow k = k^* \; .
\ee
This defines the {\it forecast mode}, that is the most probable 
prediction (7). It is this course of thinking which was accepted in Refs. 
[18--20]. However, in the real life it is not necessarily the most 
probable case that happens, because a time series is a realization of a 
random process. What we need for generalizing the approach is to be able 
to present the whole spectrum of possible scenarios weighted with the 
corresponding probabilities, which would allow us to calculate the 
statistical characteristics of the random process.

Before going to this generalization, let us make a note with regard to 
the usage of the self-similar exponentials (4). The form of the latter 
reminds us the iterated exponentials introduced by Euler [34], which have 
been studied in mathematical literature [35,36], where they are labelled 
by various names, like iterated exponentials, infinite exponentials, 
continued exponentials, multiple exponentials, stacked exponents, 
exponential towers, hypertowers, hyperexponents, superexponents, endless 
exponents, power sequences, reiterated exponentials, and so on. Except 
their form, our self-similar exponentials (4) are quite different from 
the Euler iterated exponentials [34]. The difference is, first of all, in 
the origin. The Euler exponential is the iterative solution of a 
transcendental equation, while the self-similar exponents are the outcome 
of the self-similar approximation theory [32], as applied to the 
polynomial function (2). The theory [32] prescribes the relation between 
the coefficients $a_n$ of the latter function. A specific feature of the 
self-similar exponentials is the existence of the control time $\tau_k$ 
defining a fixed point of the approximation cascade [23--25]. In general 
[32], the self-similar exponentials can have a more complicated structure 
involving noninteger powers of the variable $t$. This kind of exponentials 
with noninteger powers yields, in the first approximation, the so-called 
stretched exponentials that are often met in various applications  [37].

Let us now return to the problem of defining the scenario probabilities. 
Assume that we are interested in what happens at the time $t=k+\Dlt t$. 
For the latter, we can construct the set $\{ f_k^*(k+\Dlt t)\}$, where
$k=1,2,\ldots,N$, of the self-similar forecasts (7). We need to define the 
probability $p_k(\Dlt t)$ for the realization, at the time $t=k+\Dlt t$,
of the forecast $f_k^*(k+\Dlt t)$, which is based on the self-similar 
analysis of $k+1$ terms from the subfamily ${\bf\Phi}_k$ of the data base
$\Bbb{X}_N$. For brevity, we shall call $p_k(\Dlt t)$ the {\it 
k-scenario probability}.

The idea of defining a probability $p$ comes from statistical mechanics 
[38] where a probability $p$ can be connected with entropy $S$ by the 
relation $p\sim e^{-S}$. Another idea originates from dynamical theory 
where there exists the notion of the so-called dynamical entropy or the 
Kolmogorov-Sinai entropy rate [39,40]. The latter, for a $d$-dimensional 
dynamical system, is given by the sum
$$
h\equiv \sum_{i=1}^d\lbd_i\Theta(\lbd_i)
$$
of positive Lyapunov exponents $\lbd_i$. Since $h$ is an entropy rate, the 
entropy itself should be written as $S=hk$, where $k=1,2,\ldots$ is 
discrete time. The Kolmogorov-Sinai entropy characterizes the asymptotic 
in time behaviour of unstable trajectories.

There are two specific features of the case we are dealing with. First, 
we consider not asymptotic in time properties of a dynamical system but 
its finite-time behaviour. And second, we need to characterize not only 
unstable trajectories but all of them, stable as well as unstable. Thus, 
we generalize the Kolmogorov-Sinai entropy rate by introducing the {\it 
summary local Lyapunov exponent}
$$
\Lambda_k \equiv \sum_{i=1}^d \lbd_{ik} \; ,
$$
being the sum of {\it all local} Lyapunov exponents $\lbd_{ik}$, positive 
as well as negative. For a one-dimensional dynamical system, we have just 
one local Lyapunov exponent $\Lambda_k=\lbd_k$. The quantity 
$S_k=\Lambda_k k$ can be both negative and positive, thence it may be called
{\it dynamical quasientropy}. Retaining the relation $p_k\sim e^{-S_k}$, 
we have $p_k\sim e^{-\Lambda_k k}$. The local Lyapunov exponent can be 
expressed through the local multiplier [23--25] as
$$
\Lambda_k = \frac{1}{k} \; \ln|m_k| \; .
$$
Hence $p_k\sim |m_k|^{-1}$, which, with the normalization condition
$$
\sum_{k=1}^N p_k(\Dlt t) = 1 \; ,
$$
results in the $k$-scenario probability
\be
p_k(\Dlt t) = \frac{|m_k^*(k+\Dlt t)|^{-1}}{Z(\Dlt t)} \; , \qquad
Z(\Dlt t) \equiv \sum_{k=1}^N \frac{1}{|m_k^*(k+\Dlt t)|} \; ,
\ee
which mathematically expresses the intuitive inverse relation between 
stability and probability. The local multipliers here are defined in 
Eq. (11).

In this way, the spectrum $\{ f_k^*(k+\Dlt t)\}_{k=1}^N$ of possible 
scenarios is weighted with the scenario probabilities (14). The average 
forecast is
\be
< f(\Dlt t)> \equiv \sum_{k=1}^N p_k(\Dlt t) f_k^*(k+ \Dlt t) \; .
\ee
As for any statistical analysis, we can define the dispersion
\be
\sgm^2(\Dlt t) \equiv < f^2(\Dlt t)> - < f(\Dlt t)>^2 \; ,
\ee
the standard deviation
\be
\sgm(\Dlt t) \equiv 
\left [ < f^2(\Dlt t)> - < f(\Dlt t)>^2 \right ]^{1/2} \; ,
\ee
having for markets the meaning of volatility, and the variance coefficient
\be
\rho(\Dlt t) \equiv \frac{\sgm(\Dlt t)}{<f(\Dlt t)>} \cdot 100\% \; .
\ee
When the actually realized value $x_{N+\Dlt t}$ for the considered moment 
of time is known, one may find the percentage error of the average forecast
(15) as
\be
\ep(\Dlt t) \equiv \frac{<f(\Dlt t)> - x_{N+\Dlt t}}{|x_{N+\Dlt t}|}
\cdot 100\% \; .
\ee
If one deals with a series of examples for which the data-base order $N$ 
and the prediction time $\Dlt t$ are fixed, one may simplify the notation 
by omitting the quantities $N$ and $\Dlt t$, for instance writing
\be
< f > = < f(\Dlt t) > \qquad
(N,\; \Dlt t, \; {\rm fixed}) \; .
\ee
The described procedure of analysing time series composes the {\it 
statistical self-similar analysis}.

\section{Examples of Market Bubbles}

To illustrate the developed procedure, we select several examples of 
market time series exhibiting bubbles, which, as is mentioned in the 
Introduction, is the most difficult and most intriguing case for analysis. 
For the uniformity of consideration, we take everywhere a six-order data 
base, that is $N=5$, and for the prediction time, we set $\Dlt t=1$. For 
convenience, the results are arranged in the form of tables.

\vskip 3mm

{\it Example 1}. The dynamics of the average index of the South African 
gold mining share prices in the period of time from the second quarter of
1986 till the third quarter of 1987. The latter index is accepted as 100 
(1987, III=100). Let us make a forecast for the fourth quarter of 1987, 
comparing it with the actual value $x_6=81.64$. The data $x_n$ and the 
results for the self-similar forecasts $f_n^*(n+1)$, the related 
local multipliers $m_n^*(n+1)$, and for the corresponding probabilities are 
given in Table 1. The average forecast (15), standard dispersion (17), 
variance coefficient (18), and the error (19), respectively, are
$$
<f> =82.926\; , \qquad \sgm=2.25\; , \qquad \rho=2.71\% \; , \qquad
\ep=1.58\% \; .
$$

\vskip 3mm

{\it Example 2}. Let the USA tobacco price index (all markets) be given from 
1965 till 1970, and we predict what happens in 1971. The value for 1990 
is taken for 100 (1990=100). The results are in Table 2. Other 
characteristics are
$$
<f> =39.74\; , \qquad \sgm=2.27\; , \qquad \rho=5.71\% \; , \qquad
\ep=-4.93\% \; .
$$

\vskip 3mm

{\it Example 3}. The behaviour of the Bolivian zinc price index from 1979 
till 1984 gives us an example of a nonmonotonic growth. We make a 
forecast for 1985. The corresponding analysis is presented in Table 3, 
where the value for 1990 is taken for 100 (1990=100). We have
$$
<f> =60.35\; , \qquad \sgm=6.41\; , \qquad \rho=10.6\% \; , \qquad
\ep=-2.74\% \; .
$$

\vskip 3mm

{\it Example 4}. The average index of Spanish share prices from the 
second quarter of 1986 till the third quarter of 1987. The time of 
interest is the fourth quarter of 1987. The analysis is in Table 4, and
$$
<f> =80.479\; , \qquad \sgm=7.354\; , \qquad \rho=9.14\% \; , \qquad
\ep=9.62\% \; .
$$

\vskip 3mm

{\it Example 5}. The Indian share price index from 1969 till 1974 
(1985=100). The time of interest is 1975. The results of analysis are in 
Table 5, and
$$
<f> =33.5\; , \qquad \sgm=3.44\; , \qquad \rho=10.2\% \; , \qquad
\ep=4.8\% \; .
$$

\vskip 3mm

{\it Example 6}. The Mexican share price index from 1989 till 1994 
(1990=100). The forecasting time is 1995. The analysis is in Table 6, and
$$
<f> =365.0\; , \qquad \sgm=47.8\; , \qquad \rho=13.1\% \; , \qquad
\ep=-6.65\% \; .
$$

\vskip 3mm

{\it Example 7}. The Korean share price index from 1973 till 1978 (1985=100).
The forecasting time is 1979. The results are in Table 7, with
$$
<f> =84.8\; , \qquad \sgm=16.4\; , \qquad \rho=19.3\% \; , \qquad
\ep=-2.35\% \; .
$$

\vskip 3mm

{\it Example 8}. The UK copper price index in 1975 to 1980 (1990=100). 
The forecasting time is 1981. The analysis is in Table 8, and
$$
<f> =67.9\; , \qquad \sgm=34.1\; , \qquad \rho=50.2\% \; , \qquad
\ep=3.56\% \; .
$$

\vskip 3mm

{\it Example 9}. The Denmark industrial share price index from 1968 till 
1973 (1985=100). The time of interest is 1974. The analysis is in table 
9, and
$$
<f> =18.1\; , \qquad \sgm=11.1\; , \qquad \rho=61.3\% \; , \qquad
\ep=-5\% \; .
$$

\vskip 3mm

{\it Example 10}. The World commodity price index from 1969 to 1974 
(1990=100). The forecast time is 1975. The results are in Table 10, and
$$
<f> =63.7\; , \qquad \sgm=43.2\; , \qquad \rho=67.8\% \; , \qquad
\ep=-10.2\% \; .
$$

\vskip 3mm

{\it Example 11}. The US silver price index from 1975 to 1980 (1990=100). 
The time of interest is 1981. The analysis is in Table 11, and
$$
<f> =248.6\; , \qquad \sgm=309.6\; , \qquad \rho=125\% \; , \qquad
\ep=12.2\% \; .
$$

\vskip 3mm

{\it Example 12}. The gold price index from 1970 to 1975 (1990=100). The 
forecasting time is 1976. The results of analysis are in Table 12, and
$$
<f> =33.7\; , \qquad \sgm=39.9\; , \qquad \rho=118\% \; , \qquad
\ep=3.5\% \; .
$$

\vskip 3mm

Let us note that it is admissible to incorporate in the above analysis 
the no-change term by formally setting $f_0^*(1)\equiv\vp_{00}=x_N$ and
ascribing to the latter the multiplier $m_0^*(1)\equiv 1$. In the 
examples considered, the probability $p_0(1)$ is always small, so that 
the no-change term practically does not contribute to the averages.

In conclusion, we have generalized the self-similar analysis of time 
series [18--20] by making this approach statistical. The scenario 
probabilities are introduced. The method makes it possible to analyse the 
whole forecast spectrum by considering different outcomes characterized
by their weights. The average forecast is defined as the average fixed 
point. The latter does not need to be compulsory very close to the most 
probable forecast or to the actually realized value, although in the 
majority of cases it is so. Several examples of market time series, 
exhibiting bubbles, illustrate the approach. Since a time series is a 
realization of a random process, the bubble burst is a stochastic event 
that can be predicted only in a probabilistic way. The most that any 
forecasting theory can achieve is to define a forecast spectrum of 
possible scenarios weighted with the corresponding probabilities. But 
being able to get such a statistical analysis means to be in a position 
of using it. In this short communication we could not (and did not plan to) 
explain all technical details of the practical usage of the statistical 
self-similar analysis. This is a separate story. Our main aim here has 
been to demonstrate the principal way of constructing such a statistical 
analysis of time series.

\vskip 5mm

{\bf Acknowledgement}

\vskip 2mm

Useful advice and criticism from E.P. Yukalova are gratefully appreciated.

\newpage

\begin{center}

{\large{\bf Table Captions}}

\end{center}

{\bf Table 1}. Statistical self-similar analysis of scenarios for the 
South African gold mining share price in 1987, IV, based on the data from 
1986, II, till 1987, III.

\vskip 5mm

{\bf Table 2}. Self-similar analysis of scenarios for the 
USA tobacco price index in 1971, based on the data from 1965 till 1970.

\vskip 5mm

{\bf Table 3}. Analysis of scenarios for the Bolivian zinc
price index in 1985, with the data base from 1979 till 1984.

\vskip 5mm

{\bf Table 4}. Analisis of scenarios for the average index of Spanish
share prices in 1987, IV, with the data base from 1986, II, till 1987, III.

\vskip 5mm

{\bf Table 5}. Analysis of scenarios for the Indian share
price index in 1975, based on the data from 1969 till 1974.

\vskip 5mm

{\bf Table 6}. Scenarios for the Mexican share
price index in 1995, with the data base from 1989 till 1994.

\vskip 5mm

{\bf Table 7}. Scenarios for the Korean share
price index in 1979, based on the data from 1973 till 1978.

\vskip 5mm

{\bf Table 8}. Scenarios for the UK copper
price index in 1981, based on the data from 1975 to 1980.

\vskip 5mm

{\bf Table 9}. Scenarios for the Denmark industrial share
price index in 1974, with the data from 1968 to 1973.

\vskip 5mm

{\bf Table 10}. Scenarios for the World commodity
price index in 1975, based on the data from 1969 to 1974.

\vskip 5mm

{\bf Table 11}. Scenarios for the US silver
price index in 1981, based on the data from 1975 to 1980.

\vskip 5mm

{\bf Table 12}. Analysis of scenarios for the gold
price index in 1976, based on the data from 1970 to 1975.

\newpage

\begin{center}

{\bf Table 1}

\vskip 3mm

\begin{tabular}{|c|c|c|c|c|c|c|c|}\hline
$n$         & $0$    & $1$    & $2$    & $3$    & $4$    & $5$ & $6$  \\ \hline
$x_n$       & 52.734 & 69.141 & 82.813 & 85.938 & 93.750 & 100 & $\bf 81.640$ \\
$f_n^*(n+1)$& $--$   & 107.122& 109.355& 82.812 & 132.501& $\infty$ &  \\
$m_n^*(n+1)$& $--$   & 1      & 1      &$-4\times 10^{-4}$ & 0.233 & 
$\infty$ &  \\
$p_n(1)$ & $--$ & $4\times 10^{-4}$ & $4\times 10^{-4}$ & 0.997 & 0.002 & 
0 & \\ \hline
\end{tabular}

\vskip 1cm

{\bf Table 2}

\vskip 3mm

\begin{tabular}{|c|c|c|c|c|c|c|c|}\hline
$n$         & $0$  & $1$  & $2$  & $3$  & $4$  & $5$  & $6$       \\ \hline
$x_n$       & 33.9 & 36.8 & 37.1 & 37.9 & 39.5 & 45.9 & $\bf 41.8$ \\
$f_n^*(n+1)$& $--$ & 54.6 & 37.5 & 38.4 & 36.5 & 41.4 &            \\
$m_n^*(n+1)$& $--$ & 1    & 1    & 0.023& 0.152& $-$0.024&          \\
$p_n(1)$    & $--$ & 0.011& 0.011& 0.464 & 0.070 & 0.445 & \\ \hline
\end{tabular}

\vskip 1cm

{\bf Table 3}

\vskip 3mm

\begin{tabular}{|c|c|c|c|c|c|c|c|}\hline
$n$         & $0$  & $1$  & $2$  & $3$  & $4$  & $5$  & $6$       \\ \hline
$x_n$       & 53.5 & 53.6 & 61.2 & 58.3 & 54.6 & 68.4 & $\bf 58.7$ \\
$f_n^*(n+1)$& $--$ & 90.5 & 44.7 & 61.3 & 59.3 & 47.0 &            \\
$m_n^*(n+1)$& $--$ & 1    & 1    &$-$0.027&$-$0.620& $-$0.288&          \\
$p_n(1)$    & $--$ & 0.023& 0.023& 0.839 & 0.037 & 0.079 & \\ \hline
\end{tabular}

\vskip 1cm

{\bf Table 4}

\vskip 3mm

\begin{tabular}{|c|c|c|c|c|c|c|c|}\hline
$n$         & $0$  & $1$  & $2$  & $3$  & $4$  & $5$  & $6$       \\ \hline
$x_n$       &58.721&62.338&63.906&79.243&76.589& 100  & $\bf 73.026$ \\
$f_n^*(n+1)$& $--$ &141.147&63.083&94.207&43.978&78.060 &            \\
$m_n^*(n+1)$& $--$ & 1    & 1    &$-$0.029& 1.109& $-$0.005 &          \\
$p_n(1)$    & $--$ & 0.004& 0.004& 0.145 & 0.004 & 0.843 & \\ \hline
\end{tabular}

\newpage

{\bf Table 5}

\vskip 3mm

\begin{tabular}{|c|c|c|c|c|c|c|c|}\hline
$n$         & $0$  & $1$  & $2$  & $3$  & $4$  & $5$  & $6$       \\ \hline
$x_n$       & 30.9 & 33.4 & 32.3 & 31.8 & 35.1 & 38.8 & $\bf 31.9$ \\
$f_n^*(n+1)$& $--$ & 43.3 & 46.3 & 31.4 & 33.7 & 32.4 &            \\
$m_n^*(n+1)$& $--$ & 1    & 1    &$-$0.170&0.132& $-$0.101&          \\
$p_n(1)$    & $--$ & 0.039& 0.039& 0.232 & 0.299 & 0.390 & \\ \hline
\end{tabular}

\vskip 1cm

{\bf Table 6}

\vskip 3mm

\begin{tabular}{|c|c|c|c|c|c|c|c|}\hline
$n$         & $0$  & $1$  & $2$  & $3$  & $4$  & $5$  & $6$       \\ \hline
$x_n$       & 57.7 & 100  & 190.1& 291.3& 325.6& 442.1& $\bf 389.3$ \\
$f_n^*(n+1)$& $--$ & 666.0& 288.9& 510.0&$\infty$& 350.1 &            \\
$m_n^*(n+1)$& $--$ & 1    & 1    & 0.020&$\infty$& $-$0.002&          \\
$p_n(1)$    & $--$ & 0.002& 0.002& 0.091& 0     & 0.906 & \\ \hline
\end{tabular}

\vskip 1cm

{\bf Table 7}

\vskip 3mm

\begin{tabular}{|c|c|c|c|c|c|c|c|}\hline
$n$         & $0$  & $1$  & $2$  & $3$  & $4$  & $5$  & $6$       \\ \hline
$x_n$       & 57.6 & 56.7 & 63.0 & 77.3 & 81.7 & 103.4& $\bf 86.8$ \\
$f_n^*(n+1)$& $--$ & 139.0& 74.3 & 94.2 & 52.5 & 60.6 &            \\
$m_n^*(n+1)$& $--$ & 1    & 1    & 0.013& 0.256& $-$0.038&          \\
$p_n(1)$    & $--$ & 0.009& 0.009& 0.698& 0.035& 0.239 & \\ \hline
\end{tabular}

\vskip 1cm

{\bf Table 8}

\vskip 3mm

\begin{tabular}{|c|c|c|c|c|c|c|c|}\hline
$n$         & $0$  & $1$  & $2$  & $3$  & $4$  & $5$  & $6$       \\ \hline
$x_n$       & 46.5 & 52.7 & 49.2 & 51.3 & 74.1 & 82.1 & $\bf 65.5$ \\
$f_n^*(n+1)$& $--$ & 92.0 & 155.9& 46.5 & 59.7 & 365.9 &            \\
$m_n^*(n+1)$& $--$ & 1    & 1    &$-$0.269& 0.213& 49.371&          \\
$p_n(1)$    & $--$ & 0.096& 0.096& 0.356& 0.450& 0.002 & \\ \hline
\end{tabular}

\newpage

{\bf Table 9}

\vskip 3mm

\begin{tabular}{|c|c|c|c|c|c|c|c|}\hline
$n$         & $0$  & $1$  & $2$  & $3$  & $4$  & $5$  & $6$       \\ \hline
$x_n$       & 12   & 14   & 13   & 12   & 17   & 26   & $\bf 19$ \\
$f_n^*(n+1)$& $--$ & 49   & 25   & 12   & 15   & 454  &            \\
$m_n^*(n+1)$& $--$ & 1    & 1    &$-$0.349& 0.126& 467  &          \\
$p_n(1)$    & $--$ & 0.078& 0.078& 0.224& 0.620&$2\times 10^{-4}$ & \\ \hline
\end{tabular}

\vskip 1cm

{\bf Table 10}

\vskip 3mm

\begin{tabular}{|c|c|c|c|c|c|c|c|}\hline
$n$         & $0$  & $1$  & $2$  & $3$  & $4$  & $5$  & $6$       \\ \hline
$x_n$       & 40.4 & 39.7 & 37.1 & 42.8 & 69.2 & 83.9 & $\bf 70.2$ \\
$f_n^*(n+1)$& $--$ & 105.8& 202.6& 37.4 & 58.3 & 37.1 &            \\
$m_n^*(n+1)$& $--$ & 1    & 1    &$-$0.320& 0.171& $-$0.392&          \\
$p_n(1)$    & $--$ & 0.074& 0.074& 0.231& 0.432& 0.189 & \\ \hline
\end{tabular}

\vskip 1cm

{\bf Table 11}

\vskip 3mm

\begin{tabular}{|c|c|c|c|c|c|c|c|}\hline
$n$         & $0$  & $1$  & $2$  & $3$  & $4$  & $5$  & $6$       \\ \hline
$x_n$       & 91.7 & 90.3 & 95.9 & 112.1& 230.1& 426.9& $\bf 218.3$ \\
$f_n^*(n+1)$& $--$ & 1273 & 918.6& 94.5 & 182.4& 76.0 &            \\
$m_n^*(n+1)$& $--$ & 1    & 1    &$-$0.149& 0.116& $-$1.93&          \\
$p_n(1)$    & $--$ & 0.056& 0.056& 0.376& 0.483& 0.029 & \\ \hline
\end{tabular}

\vskip 1cm

{\bf Table 12}

\vskip 3mm

\begin{tabular}{|c|c|c|c|c|c|c|c|}\hline
$n$         & $0$  & $1$  & $2$  & $3$  & $4$  & $5$  & $6$       \\ \hline
$x_n$       & 9.4  & 10.6 & 15.2 & 25.4 & 41.5 & 42.0 & $\bf 32.5$ \\
$f_n^*(n+1)$& $--$ & 42.5 & 127.5&$\infty$& 84.0 & 9.3 &            \\
$m_n^*(n+1)$& $--$ & 1    & 1    &$\infty$& 1.811& $-$0.176 &       \\
$p_n(1)$    & $--$ & 0.121& 0.121& 0      & 0.067  & 0.690 & \\ \hline
\end{tabular}

\end{center}

\end{document}